\documentstyle[aps,epsfig,multicol,subfigure]{revtex}
\begin{document}
\draft


\title{Maximal entanglement from quantum random walks}

\author{B. All\'es$^{\rm a}$, S. G\"und\"u\c c$^{\rm b}$, Y. G\"und\"u\c c$^{\rm c}$}

\address{$^{\rm a}$INFN Sezione di Pisa, Pisa, Italy}
\address{$^{\rm b}$Hacettepe Vocational School, Hacettepe University, 06800 Beytepe, Ankara, Turkey}
\address{$^{\rm c}$Department of Physics, Hacettepe University, 06800 Beytepe, Ankara, Turkey}

\maketitle

\begin{abstract}
The conditions under which entanglement becomes maximal are sought in the general one--dimensional quantum random walk
with two walkers. Moreover, a one--dimensional shift operator for the two walkers is introduced and its performance
in generating entanglement is analyzed as a function of several free parameters, some of them coming from the shift
operator itself and some others from the coin operator. To simplify the investigation an averaged entanglement is defined.
\end{abstract}

\begin{multicols}{2}


\section{Introduction}
\label{section1}

Quantum random walks~\cite{aharonov,kempe,carneiro,kendon} (QRW) are a generalization of classical random walks.
At each step of
the latter the position of a classical particle (walker) is shifted according to the result of tossing a
coin. Thus, the position of the walker is decided by following a probability distribution over classically
well--defined positions. In the QRW case the tossing of the coin is substituted by the action of a unitary
operator (that shall be called coin operator) on a 2--state system (called coin state) and the motion of
the walker by a unitary transformation of the position state by a shift operator. The state of the walker
is in general a quantum superposition of several position eigenstates.

Many aspects of such systems have been studied in the past~\cite{konno,abal,feynman,meyer}. In~\cite{grimmett} the
consequences of making one walker move in more than one dimension are analyzed. The interplay between decoherence and
entanglement has been unveiled in~\cite{maloyer} while in~\cite{liu} two coins have been used. The so--called meeting
problem has been studied in~\cite{stefanak} for two walkers. We end this short review of recent results by recalling
Ref.~\cite{romanelli} where the evolution of the chirality of the coin has been investigated.

QRW also attract interest in some specific research topics. For example, the efficiency of the energy transfer
in photosynthesis can be raised at the level of about 99\% if it is modelled by a QRW interacting with the
thermal fluctuations of the environment~\cite{mohseni}. Besides, quantum transport properties of electron systems and
dielectric breakdown driven by strong electric fields can also be studied by utilizing QRW~\cite{oka}. In ancilla--based
quantum computation schemes~\cite{ancilla} the ancilla plays the r\^ole of the coin and the measurement produces
entanglement between the qubits. Moreover, experimental set--ups for QRW are nowadays available~\cite{du,ryan,perets}.

In the present paper we propose a quantum walk made of one spin $\frac{1}{2}$ state (the coin) coupled to two particles
(the walkers) whose positions can take any integer value on an infinite line. QRW with two walkers are a valid tool
for obtaining highly entangled states~\cite{venegas,goyal}. Indeed, performing measurements on the 2--state coin system
after $n$ random walk steps may yield position states with strong entanglement among the two walkers. In particular, our
QRW has been modelled in such a way that a further position measurement on one walker fixes the position of the other,
thus rendering the correlation among the two walkers a maximum.

The relevance of entanglement in QRW is crucial. While this property does not appear in classical walks, it becomes the
characterizing ingredient of QRW. The interest of entanglement lies in the fact that it is of utmost relevance for research
in quantum computation protocols and quantum information~\cite{farhi,nayak,ambainis,watrous,nielsen} (cryptography, communication,
algorithms, etc.). We want to study the growth of entanglement among the two walkers and, in particular, to discover whether maximal
entanglement may be achieved. We also pay attention to the quantum probability $P$ to obtain such entangled states after
the coin measurement. 

As a novel ingredient of the present paper, we will introduce a shift operator that to some extent
generalizes the ones so far used in literature. Its form contains several free parameters which, added to the parameters
of the coin operator, supply enough freedom to easily find conditions under which entanglement is maximally enhanced.

The plan of the paper is as follows. In section~\ref{section2} the model for QRW is presented in detail. It includes a
discussion about the general type of shift operator. An analytic study of the QRW during the first steps in the walk is
shown in section~\ref{section3}. This approach, though applicable to a very limited number of random walk steps,
turns out to be extremely useful to interpret the numerical results derived in section~\ref{section4}.
Indeed, in the latter section, thanks to a numerical treatment of the random walk evolution, we study the
entanglement at an arbitrarily large number of steps. In this section an averaged entanglement will be
defined to simplify the search for the conditions of maximal entanglement. Some conclusive comments will
be given in section~\ref{section5}.

\section{The spin--particle QRW}
\label{section2}

We begin by stating the precise definition of the model under study. The two walkers can roam on a line
of discrete spatial positions. The accessible position eigenstates are $\vert i,i\rangle$ for $i\in{\bf Z}$ ($i=0$ is
the origin) where the first item refers to the position of the first walker and the second one of the second walker.
Thus the two walkers are supposed to stay together during the walk.
The coin state is a spin $\frac{1}{2}$ system. Its eigenstates are the spin component along the $Z$--axis.
The complete quantum state is a sum of terms of the form $\vert\phi\rangle\equiv\vert s\rangle\otimes\vert\psi\rangle$
where the first factor refers to the spin and the second one indicates the position state. The $\otimes$ symbol
emphasizes the tensor structure of the state space. The most general pure position state contemplated by our model is
$\vert\psi\rangle=\sum_i c_i \vert i,i\rangle$ with $\sum_i \vert c_i\vert^2=1$. The eigenvectors of the
$Z$--component of the spin are denoted by an
arrow: $\vert\uparrow\rangle$ ($\vert\downarrow\rangle$) stands for the $Z$--component $+\frac{1}{2}$ ($-\frac{1}{2}$).
Its most general pure state is $d_\uparrow\vert \uparrow\rangle + d_\downarrow\vert \downarrow\rangle$ with
$\vert d_\uparrow\vert^2+\vert d_\downarrow\vert^2=1$. A similar model was introduced in~\cite{venegas}.

The above model is hardly realizable in laboratory (for a more physical model with two walkers see for example~\cite{ref}
and references therein). Rather it must be viewed as the limiting case where entanglement is expected to be maximal. Indeed,
imagine a more general QRW model for which the position state of the two walkers after the spin measurement (performed
after $n$ QRW steps) is $\sum_{i,j} c_{ij}\vert i,j\rangle$ ($i,j\in{\bf Z}$ and $\sum_{i,j}\vert c_{ij}\vert^2=1$) where,
as before, the first (second) item in the ket refers to the first (second) walker. This state displays large entanglement
among the walkers if after a further measurement, this time on the position state of one of the two walkers, the position
of the other walker gets fixed with maximum probability. Evidently, the limiting case of the above circumstance occurs when
$j$ is a function of $i$, $j=f(i)$ where $f$ is a predetermined known function. As for $f$ we have chosen the identity function.
Moreover, since we have chosen such an idealized QRW model, the precise connection among the spin variable and the walkers
will not be better specified. In the rest of the paper we will study the entropy that entangles the two particles in the
state $\sum_{i} c_{i}\vert i,i\rangle$ after the coin measurement.

Since we want to study the generation of entanglement, throughout the paper the factor corresponding to the position of
the walkers in the initial state $\vert\phi\rangle_{0}$ of the QRW shall be free of entanglement,
\begin{equation}
\vert\phi\rangle_{0}=\left(d_\uparrow\vert \uparrow\rangle + d_\downarrow\vert \downarrow\rangle\right)
\otimes\vert 0,0\rangle,
\label{stateinitial}
\end{equation}
where the above normalization condition regarding coefficients $d_{\uparrow},d_{\downarrow}$ is assumed and the subscript~0 stands
for initial state (0--th step of the QRW). Furthermore, since for any spin state $\vert s\rangle$ there exists some axis $Z^\prime$
such that $\vert s\rangle=\vert\uparrow^\prime\rangle$, the state in~(\ref{stateinitial}) can be simplified to (dropping primes)
\begin{equation}
\vert\phi\rangle_{0}=\vert\uparrow\rangle\otimes\vert0,0\rangle ,
\label{stateinitial2}
\end{equation}
by adequately choosing the reference axes. Such a simplification does not entail a loss of generality
for the reason that the position state and the coin state in~(\ref{stateinitial}) are uncorrelated.

A single step in the random walk consists in the concatenation of two operators,
$U_{\rm shift}\cdot U_{\rm coin}$, iteratively applied to $\vert\phi\rangle$.
Operator $U_{\rm shift}$ acts on position states and $U_{\rm coin}$ on spin states.

Let us begin by introducing the shift operator $U_{\rm shift}$. A quite general form is
\begin{eqnarray}
U_{\rm shift}
&=&\sum_n\alpha^{(n)}_{\uparrow\uparrow}\vert\uparrow\rangle\langle\uparrow\vert\sum_i\vert i+n,i+n\rangle\langle i,i\vert
\nonumber\\
&+&\sum_n\alpha^{(n)}_{\uparrow\downarrow}\vert\uparrow\rangle\langle\downarrow\vert\sum_i\vert i+n,i+n\rangle\langle i,i\vert
\nonumber\\
&+&\sum_n\alpha^{(n)}_{\downarrow\uparrow}\vert\downarrow\rangle\langle\uparrow\vert\sum_i\vert i+n,i+n\rangle\langle i,i\vert
\nonumber\\
&+&\sum_n\alpha^{(n)}_{\downarrow\downarrow}\vert\downarrow\rangle\langle\downarrow\vert\sum_i\vert i+n,i+n\rangle\langle i,i\vert,
\end{eqnarray}
with $\alpha^{(n)}_{ss^\prime}$ complex coefficients. They must satisfy several constraints derived from
imposing unitarity, $U_{\rm shift}U_{\rm shift}^\dagger={\bf I}$, (${\bf I}$ is the unit operator).
After some algebra, the l.h.s. of this condition is
\begin{eqnarray}
\sum_{n,m,i}&\Big[&\vert\downarrow\rangle\langle\uparrow\vert\,\vert i+n,i+n\rangle\langle i+m,i+m\vert\nonumber\\
                  &&\hspace{1.3cm}\times\left(\alpha^{(n)}_{\downarrow\uparrow}\alpha^{(m)\ast}_{\uparrow\uparrow}+
                        \alpha^{(n)}_{\downarrow\downarrow}\alpha^{(m)\ast}_{\uparrow\downarrow}\right)\nonumber\\
&+&\vert\uparrow\rangle\langle\downarrow\vert\,\vert i+m,i+m\rangle\langle i+n,i+n\vert\nonumber\\
                  &&\hspace{1.3cm}\times\left(\alpha^{(n)\ast}_{\downarrow\uparrow}\alpha^{(m)}_{\uparrow\uparrow}+
                        \alpha^{(n)\ast}_{\downarrow\downarrow}\alpha^{(m)}_{\uparrow\downarrow}\right)\nonumber\\
&+&\vert\uparrow\rangle\langle\uparrow\vert\,\vert i+n,i+n\rangle\langle i+m,i+m\vert\nonumber\\
                  &&\hspace{1.3cm}\times\left(\alpha^{(n)}_{\uparrow\uparrow}\alpha^{(m)\ast}_{\uparrow\uparrow}+
                        \alpha^{(n)}_{\uparrow\downarrow}\alpha^{(m)\ast}_{\uparrow\downarrow}\right)\nonumber\\
&+&\vert\downarrow\rangle\langle\downarrow\vert\,\vert i+n,i+n\rangle\langle i+m,i+m\vert\nonumber\\
                  &&\hspace{1.3cm}\times\left(\alpha^{(n)}_{\downarrow\downarrow}\alpha^{(m)\ast}_{\downarrow\downarrow}+
                        \alpha^{(n)}_{\downarrow\uparrow}\alpha^{(m)\ast}_{\downarrow\uparrow}\right)\Big],
\label{sumsvil}
\end{eqnarray}
which must equal
\begin{equation}
{\bf I}=\Big(\vert\uparrow\rangle\langle\uparrow\vert+\vert\downarrow\rangle\langle\downarrow\vert\Big)
\sum_i\vert i,i\rangle\langle i,i\vert .
\end{equation}
To fulfil the condition, the first two terms in~(\ref{sumsvil}) must be zero and note also that the second term is the
h.c. of the first one. These considerations suggest to define the complex 2--vectors
$V_1^{(n)}\equiv(\alpha^{(n)}_{\uparrow\uparrow},\alpha^{(n)}_{\uparrow\downarrow})$ and
$V_2^{(n)}\equiv(\alpha^{(n)}_{\downarrow\uparrow},\alpha^{(n)}_{\downarrow\downarrow})$ because in terms of them
the above conditions become $V_1^{(n)}\cdot V_2^{(m)\ast}=0$, $V_1^{(n)}\cdot V_1^{(m)\ast}=0$,
$V_2^{(n)}\cdot V_2^{(m)\ast}=0$ for all $n\not=m$ and $V_1^{(n)}\cdot V_2^{(n)\ast}=0$,
$V_1^{(n)}\cdot V_1^{(n)\ast}=V_2^{(n)}\cdot V_2^{(n)\ast}=1$ for all $n$.
Since in complex 2--space there are only two independent orthonormal vectors, there is no much freedom to choose
$V_1^{(n)}$ and $V_2^{(n)}$. The most general solution reduces the sum over $n$ and $m$ to one single term,
$n=p$, $m=q$ for $p,q$ fixed integers and $V_1^{(p)}=(\alpha,\beta)$, $V_2^{(q)}=(-\beta^\ast,\alpha^\ast)$ (for
complex numbers $\alpha$, $\beta$) and all the other $V$--vectors equal to zero. Moreover the two coefficients
$\alpha,\beta$ satisfy $\vert\alpha\vert^2+\vert\beta\vert^2=1$. Thus, the shift operator is
\begin{eqnarray}
U_{\rm shift}&=&\Big(\alpha\vert\uparrow\rangle\langle\uparrow\vert+\beta\vert\uparrow\rangle\langle\downarrow\vert\Big)
               \sum_i\vert i+p,i+p\rangle\langle i,i\vert\nonumber\\
&\hspace{-0.4cm}+&\hspace{-0.2cm}\Big(\alpha^\ast\vert\downarrow\rangle\langle\downarrow\vert-\beta^\ast\vert\downarrow\rangle\langle\uparrow\vert\Big)
               \sum_i\vert i+q,i+q\rangle\langle i,i\vert,
\label{shiftdefinitivo}
\end{eqnarray}
with $p,q\in{\bf Z}$ fixed and $\alpha,\beta$ verifying the above normalization.

For $p=q$ no entanglement is generated by the QRW as every term in $U_{\rm shift}$ would move
the couple of particles to the same spatial position. Instead, entanglement appears
as far as $p\not=q$, the specific values of $p,q$ being immaterial since
different pairs $p,q$ are related by rescalings on the line of discrete positions where the walkers wander.
During the present study we will stick to the values $p=+1$ and $q=-1$ all the time.

Without loss of generality, we can take $\alpha$ real in~(\ref{shiftdefinitivo}). Indeed, the spin eigenstates
$\vert\uparrow\rangle,\vert\downarrow\rangle$ can be redefined in such a way to absorb the phase of $\alpha$.
Calling $\zeta$ such a phase, transformations $\vert\uparrow\rangle\to e^{{\rm i}\zeta/2}\vert\uparrow\rangle$ and
$\vert\downarrow\rangle\to e^{-{\rm i}\zeta/2}\vert\downarrow\rangle$ eliminate it from~(\ref{shiftdefinitivo}).
There is no similar procedure able to withdraw the phase of $\beta$ too. Thus, the $U_{\rm shift}$ operator depends on
only two variables: $\vert\alpha\vert$ (henceforth called just $\alpha>0$) and arg$(\beta)$. In literature the choice
$\alpha=1$, $\beta=0$ is generally found.

The coin operator $U_{\rm coin}$ can be represented by a unitary $2\times2$ matrix. The most general such a matrix is an
element of the U(2) group,
\begin{eqnarray}
U_{\rm coin}&=&\left(\begin{array}{cc}
              \sqrt{\rho}  & \sqrt{1-\rho}\; e^{{\rm i}(\theta-\eta)}\\
             -\sqrt{1-\rho}\; e^{-{\rm i}(\theta+\eta)} & \sqrt{\rho}\; e^{-2{\rm i}\eta}\\
  \end{array}\right)\,e^{{\rm i}\varphi},\nonumber\\
&&
\label{analytic.1}
\end{eqnarray}
where $0\leq\theta,\eta\leq\pi$ and $0\leq\varphi<2\pi$ are arbitrary phases and $0\leq\rho\leq 1$.
When $\varphi=\eta$ (\ref{analytic.1}) becomes an element of the subgroup SU(2). Actually,
the overall phase $\varphi$ plays no r\^ole in the following discussion, so in practice we could limit ourselves to
study SU(2) matrices to represent the most general coin operator.

Particular instances of the unitary matrix $U_{\rm coin}$ are the Hadamard coin ($\rho=1/2$, $\varphi=0$ and $\eta=\theta=\pi/2$),
\begin{eqnarray}
U_H=\frac{1}{\sqrt{2}}\left(\begin{array}{cc}
        +1  & +1 \\
        +1  & -1 \\
  \end{array}\right),
\label{analytic.2}
\end{eqnarray}
the Kempe coin~\cite{kempe} ($\rho=1/2$, $\varphi=\eta=0$ and $\theta=\pi/2$),
\begin{eqnarray}
U_K=\frac{1}{\sqrt{2}}\left(\begin{array}{cc}
        +1  & +{\rm i} \\
        +{\rm i}  & +1 \\
  \end{array}\right),
\label{analytic.kempe}
\end{eqnarray}
or the $Z$ coin ($\rho=1$, $\eta=\pi/2$, $\varphi=0$ and $\theta$ arbitrary),
\begin{eqnarray}
U_Z=\left(\begin{array}{cc}
        +1  & 0 \\
        0  & -1 \\
  \end{array}\right).
\label{analytic.3}
\end{eqnarray}

At any moment the walk can be stopped and the value of the spin measured. If the measurement was
performed after each step in the iteration then a classical random walk would result. Hence, our
interest for QRW consists in relegating the measure after many steps because in this fashion clear
entanglement is displayed. Let us consider a measurement performed only after $n$ QRW steps giving spin
$s$ and a position pure state, result of the wavefunction collapse, denoted by $\vert\psi\rangle_n^s$ and equal to
\begin{equation}
\vert\psi\rangle_n^s\equiv\sum_i c_i\vert i,i\rangle.
\label{psicollap}
\end{equation}
If the spin measurement yields up (down) spin we will write
$\vert\psi\rangle_n^{\rm up}$ ($\vert\psi\rangle_n^{\rm down}$). The evident entanglement created among the two
walkers associated with $\vert\psi\rangle_n^s$ shall be calculated by the von Neumann entropy
\begin{equation}
{E}_n\equiv-\sum_i \vert c_i\vert^2\log_2 \vert c_i\vert^2 ,
\label{entanglementdef}
\end{equation}
where the subscript indicates that the measurement has been performed after the $n$--th step. Certainly
$E_n\geq0$ and maximal entanglement is attained when the values $\vert c_i\vert$ are all equal. So, if the sum
in $\vert\psi\rangle^s_n=\sum_i c_i\vert i,i\rangle$ contains $N$ terms, $(E_n)_{\rm max}=\log_2 N$.
We shall usually calculate the normalized entanglement defined as
\begin{equation}
{\cal E}_n\equiv\frac{E_n}{(E_n)_{\rm max}}.
\label{normentanglement}
\end{equation}
This choice of units allows to easily understand when maximal entanglement has been attained.
As the number $N$ of terms in $\vert\psi\rangle_n^s$ in general depends on the step $n$, also the denominator
in~(\ref{normentanglement}) depends on $n$. When $N=1$ both $E_n$ and ${\cal E}_n$ vanish, the former due
to~(\ref{entanglementdef}) and the latter by definition.

\section{Analytic study of entanglement generation}
\label{section3}

The first few steps of a QRW can be followed analytically. Starting from state~(\ref{stateinitial2}), using the coin
operator~(\ref{analytic.1}) (where the irrelevant global phase $e^{{\rm i}\varphi}$ has been skipped) and the shift
operator~(\ref{shiftdefinitivo}) it is easy to see that entanglement can only arise after two steps.
Indeed, the walkers plus coin state after the first step (indicated with a subscript~1) is
\begin{eqnarray}
\vert\phi\rangle_1&=&(\alpha\sqrt{\rho}-\beta\sqrt{1-\rho}\,e^{-{\rm i}(\theta+\eta)})\vert\uparrow\rangle\otimes\vert1,1\rangle\nonumber\\
 -(\beta^\ast\sqrt{\rho}&+&\alpha\sqrt{1-\rho}\,e^{-{\rm i}(\theta+\eta)})\vert\downarrow\rangle\otimes\vert-1,-1\rangle,
\label{phi1}
\end{eqnarray}
and the wavefunctions after the measurement, $\vert\psi\rangle^{\rm up}_1=\vert1,1\rangle$ and $\vert\psi\rangle^{\rm down}_1=\vert-1,-1\rangle$,
whenever they exist, are clearly free of entanglement.
Let us assume that the spin measurement after the second step yielded $+\frac{1}{2}$. The resulting entangled position state is
\begin{eqnarray}
\vert\psi\rangle^{\rm up}_2&\propto&\left(\alpha\sqrt{\rho}-\beta\sqrt{1-\rho}\;e^{-{\rm i}(\theta+\eta)}\right)^2\vert2,2\rangle\nonumber\\
                &-&e^{-2{\rm i}\eta}\left\vert\alpha\sqrt{1-\rho}+\beta\sqrt{\rho}\;e^{-{\rm i}(\theta+\eta)}\right\vert^2\vert0,0\rangle .
\label{condition_upstate}
\end{eqnarray}
The presence of the proportionality symbol reminds the lack of a normalization constant. We will often write the wavefunction
resulting from the spin measurement in this way purposely because it renders more transparent the evaluation of the probability
$P$ with which such a state comes out ($P$ is just the square of the missing normalization). It is easy
to verify that maximal entanglement is achieved when the condition
\begin{eqnarray}
&&\left\vert\alpha\sqrt{\rho}-\beta\sqrt{1-\rho}\;e^{-{\rm i}(\theta+\eta)}\right\vert\nonumber\\
&&=\left\vert\alpha\sqrt{1-\rho}+\beta\sqrt{\rho}\;e^{-{\rm i}(\theta+\eta)}\right\vert
\label{condition_up}
\end{eqnarray}
holds. If the Hadamard coin were used, the condition would became $\vert\alpha+\beta\vert=\vert\alpha-\beta\vert$ which is
fulfilled whenever $\alpha$ is real and $\beta$ pure imaginary. For the Kempe coin, condition~(\ref{condition_up})
reduces to $\vert\alpha+{\rm i}\beta\vert=\vert\alpha-{\rm i}\beta\vert$ which is satisfied for any $\alpha$ and $\beta$ as
far as both are real. For the $Z$ coin, condition~(\ref{condition_up}) requires $\vert\alpha\vert=\vert\beta\vert$.

If the result of the measurement after two steps yields spin $-\frac{1}{2}$, then the position state is
\begin{eqnarray}
\vert\psi\rangle^{\rm down}_2\hspace{-0.2mm}&\propto&\hspace{-0.2mm}\Big(\alpha\beta(1-\rho)e^{-2{\rm i}(\theta+\eta)}+
      \vert\beta\vert^2\sqrt{\rho(1-\rho)}e^{-{\rm i}(\theta+\eta)}\nonumber\\
   &&-\alpha^2\sqrt{\rho(1-\rho)}e^{-{\rm i}(\theta+\eta)}-\alpha\beta^\ast\rho\Big)\vert0,0\rangle\nonumber\\
   &+&\Big(\alpha\beta^\ast(1-\rho)e^{-2{\rm i}\eta}+(\beta^\ast)^2\sqrt{\rho(1-\rho)}e^{{\rm i}(\theta-\eta)}\nonumber\\
   &&-\alpha^2\sqrt{\rho(1-\rho)}e^{-{\rm i}(\theta+\eta)}e^{-2{\rm i}\eta}\nonumber\\
   &&-\alpha\beta^\ast\rho e^{-2{\rm i}\eta}\Big)\vert-2,-2\rangle.
\label{condition_down}
\end{eqnarray}
To find the circumstances under which entanglement becomes a maximum we discuss different values of $\rho$ separately.
When either $\rho=0$ or~1 then only one term survives in each coefficient of~(\ref{condition_down}) and maximal entanglement
occurs for all $\theta$, $\eta$, $\alpha$ and arg$(\beta)$. For $\rho=\frac{1}{2}$ entanglement is a maximum again
for all possible values that the parameters can take. To see this we substitute $\rho=\frac{1}{2}$
in~(\ref{condition_down}) and obtain after some algebra
\begin{eqnarray}
\vert\psi\rangle^{\rm down}_2&\propto&e^{-{\rm i}(\theta+\eta)}\Big(\vert\beta\vert^2-\alpha^2+
                                     2{\rm i}\alpha\vert\beta\vert\sin\Delta\Big)\vert0,0\rangle\nonumber\\
&&+e^{-2{\rm i}\eta}e^{-{\rm i}\,{\rm arg}(\beta)}\Big((2\vert\beta\vert^2-1)\cos\Delta\nonumber\\
&& -{\rm i}\sin\Delta\Big)\vert-2,-2\rangle,
\end{eqnarray}
where $\Delta\equiv{\rm arg}(\beta)-\theta-\eta$. It is easy to verify that both coefficients have the same modulus.

An instance of maximum entanglement can also be deduced for the general case $0\leq\rho\leq1$. First note that every term in
the two coefficients in~(\ref{condition_down}) are equal, apart from phases. Imposing that the phases are also equal termwise
(except for a possible global phase), we discover that maximal entanglement occurs when arg$(\beta)=\theta+\eta$.

A generalization of these analytic results to $n$ iterations of the QRW is too complicated as the length of the coefficients
in $\vert\psi\rangle^{\rm up,down}_n$ grows exponentially with $n$. Then, in the following section we study the above problem numerically
with a computer program. The program yields exact numerical results but without the corresponding explicit analytical expression.

\section{Entanglement after an arbitrary number of steps}
\label{section4}

In this section the entanglement created by measuring the spin of the coin after an arbitrary number of
QRW steps having started with the initial state~(\ref{stateinitial2}) shall be studied. To this end we introduce an
averaged entanglement $\overline{\cal E}_n$ defined over the first $n$ steps as follows. Consider $n$ replicas of the
QRW and evolve each of them independently of the others. In the first replica we measure the spin and then the
entanglement after the first step obtaining ${\cal E}_1$, in the second one after the second step obtaining ${\cal E}_2$,
etc., thus gathering the collection ${\cal E}_1$, ${\cal E}_2$, ${\cal E}_3$, $\dots$, ${\cal E}_n$. On the understanding
that each measurement yields the same result ({\it i.e.}, either all measurement outputs give up spin or all give down spin), we define
\begin{equation}
\overline{\cal E}_n\equiv\frac{1}{n-1}\sum_{a=2}^n{\cal E}_a .
\label{averageentanglement}
\end{equation}
The first step ($a=1$) is excluded from the sum in the average because it generates zero entanglement
in any case. Note that in~(\ref{averageentanglement}) the subscript $n$ does not mean the step after which
the measurement is done, as in~(\ref{entanglementdef}), but the number of steps over which the average is performed.
The convenience of using such an averaged entanglement becomes obvious when it is~1 because,
being the mean of many positive numbers not larger than~1, such a result for $\overline{\cal E}_n$
necessarily implies that all of them are precisely~1, {\it i.e.} it identifies situations where entanglement is
constantly maximal (therefore the precise value of $n$ used to evaluate $\overline{\cal E}_n$ is largely immaterial,
as long as it is sufficiently large). This assertion applies equally well when $\overline{\cal E}_n=0$, but clearly
this case is physically less interesting.

A computer code was prepared to implement the above--described procedure for the QRW during a number $n$ of steps. It
calculates step by step the entanglement, either averaged or not, as well as the probability of having up or down spin after the measurement
of the coin state and also the number of terms $N$ in the collapsed position wavefunction (the latter enables us to compute the
exact value of $(E_n)_{\rm max}$). The numerical procedure used to extract $N$ was demanding that $\vert c_i\vert$ in~(\ref{psicollap}) be larger
than a pre--fixed threshold, in our case $10^{-10}$. All the results presented in this section have been obtained by using this code.

The next three subsections contain the study for the three coin operators of section~\ref{section2}, namely the Hadamard $U_H$,
the Kempe $U_K$ and $U_Z$, respectively. The last subsection is dedicated to the general coin case, (expression~(\ref{analytic.1})
with $\varphi=0$).

\vskip 1cm

\begin{figure}
\includegraphics[scale=0.33]{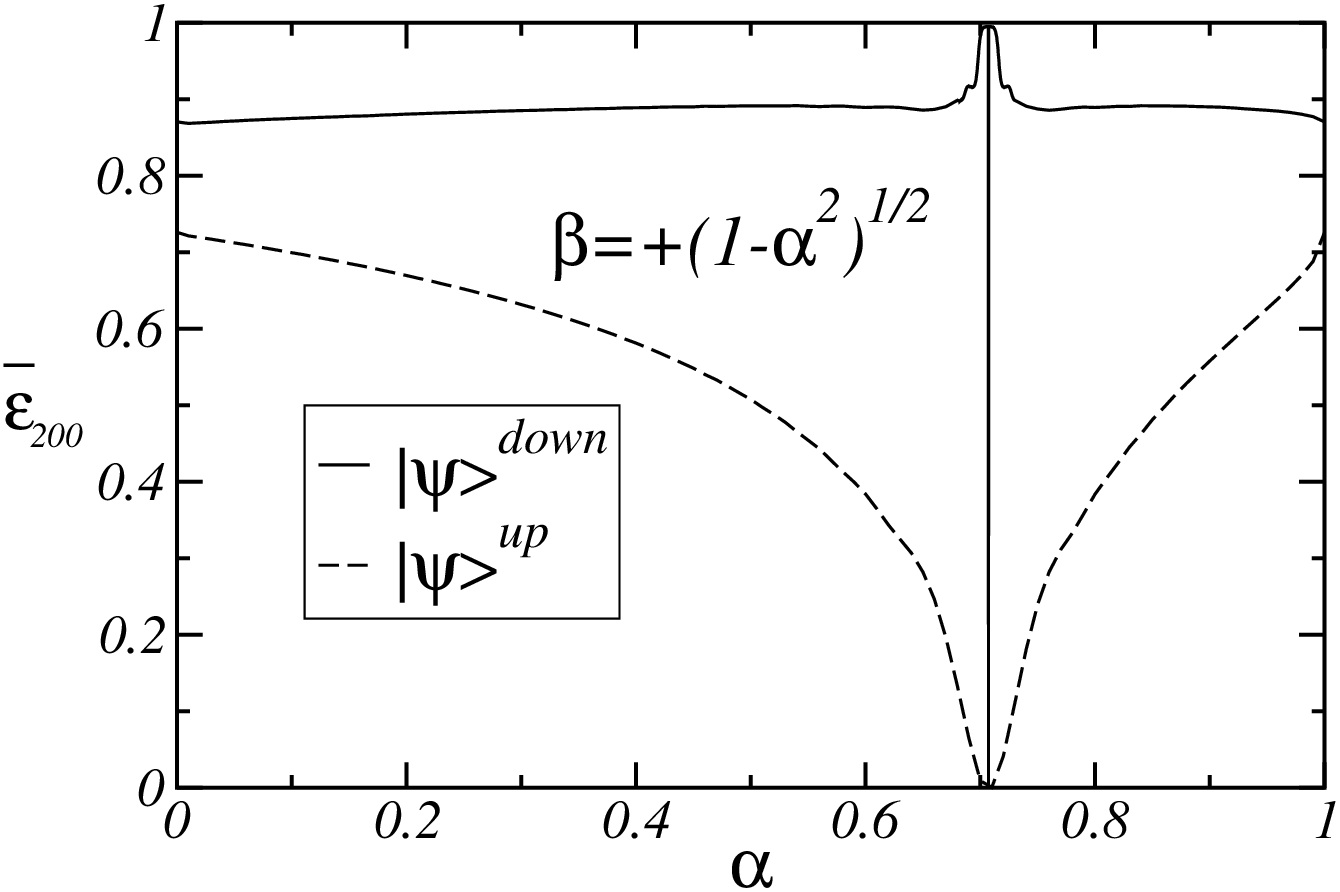}
\caption{Averaged entanglement obtained during the first 200 steps of a QRW with the Hadamard coin~(\ref{analytic.2}) and shift
operator~(\ref{shiftdefinitivo}) with $\alpha,\beta$ real. The plots for down (up) spin output of the measurements are shown with a
continuous (dashed) line. Symbols $\vert\psi\rangle^{\rm up,down}$ here and in the next figures refer to the output of the spin measurement.}
\label{Fig1}
\end{figure}

\subsection{Hadamard coin}
\label{section4.H}

This is the most frequently used coin operator for the kind of problems at hand. Its expression is given in~(\ref{analytic.2}).
To study the entanglement generated by a QRW evolved with this coin, we first calculate $\overline{\cal E}_n$ for $n=200$ steps for
real and positive values of $\alpha$ and $\beta=+\sqrt{1-\alpha^2}$ in~(\ref{shiftdefinitivo}). The result is displayed in
Fig.~\ref{Fig1} as a function of $\alpha$ ($0\leq\alpha\leq1$).

The plots in Fig.~\ref{Fig1} show that as $\alpha$ and $\beta$ tend to be alike and real (this means both approach the value
$\frac{1}{\sqrt{2}}$) entanglement tends to be maximal (minimal) if the measurement yields down (up) spin. The number of entangled
terms $N$ in the collapsed wavefunction after the measurement is equal to the step number $n$. This must be interpreted as that entanglement
gets richer and richer as the QRW goes on. However at the exact values $\alpha=\beta=\frac{1}{\sqrt{2}}$ entanglement suddenly drops
to zero for both kinds of measurement outputs. This is how the vertical line in Fig.~\ref{Fig1} (which applies to the down spin measurement)
must be deciphered. Moreover, the probability of having such maximally entangled states (for
$\alpha\approx\beta$ and both real) after the measurement is extremely small, $P\ll1$. 

The origin of all the above facts can be clarified by using the analytical expressions obtained in section~\ref{section3}.
First of all we recall that for any values of $\alpha$ and $\beta$ the first step always gives no entanglement at all. For instance, taking
$\alpha=\beta=\frac{1}{\sqrt{2}}$ the state after one step of QRW is $\vert\phi\rangle_1=\vert\uparrow\rangle\otimes\vert1,1\rangle$.
Proceeding with the choice
$\alpha=\beta=\frac{1}{\sqrt{2}}$, a calculation similar to that exhibited in section~\ref{section3} allows to show that
$\vert\phi\rangle_2=\vert\uparrow\rangle\otimes\vert2,2\rangle$. By induction this can be immediately generalized to any step, getting
$\vert\phi\rangle_n=\vert\uparrow\rangle\otimes\vert n,n\rangle$. Hence, entanglement vanishes when $\alpha=\beta=\frac{1}{\sqrt{2}}$.

Now we show why for step $n=2$ the entanglement of the state resulting from a down spin measurement approaches its maximal value
but suddenly drops to zero at $\alpha=\beta=\frac{1}{\sqrt{2}}$. Taking $\alpha$,$\beta$ in~(\ref{condition_down}) real and close to each
other leads to the expression $\vert\psi\rangle^{\rm down}_2\propto(\beta^2-\alpha^2)(\vert-2,-2\rangle-\vert0,0\rangle)$ which indeed
displays maximal entanglement although it suddenly becomes zero as soon as $\alpha$ exactly equals $\beta$. This expression also indicates
that despite the entanglement of the state resulting from a down spin measurement converges to the maximal value, as $\alpha$ tends to $\beta$
the probability of obtaining such a measurement output becomes very small, of order $O(\vert\beta^2-\alpha^2\vert^2)$. It seems
reasonable to expect that an analogous mechanism explains the low probability also for $n>2$.

\begin{figure}
\includegraphics[scale=0.33]{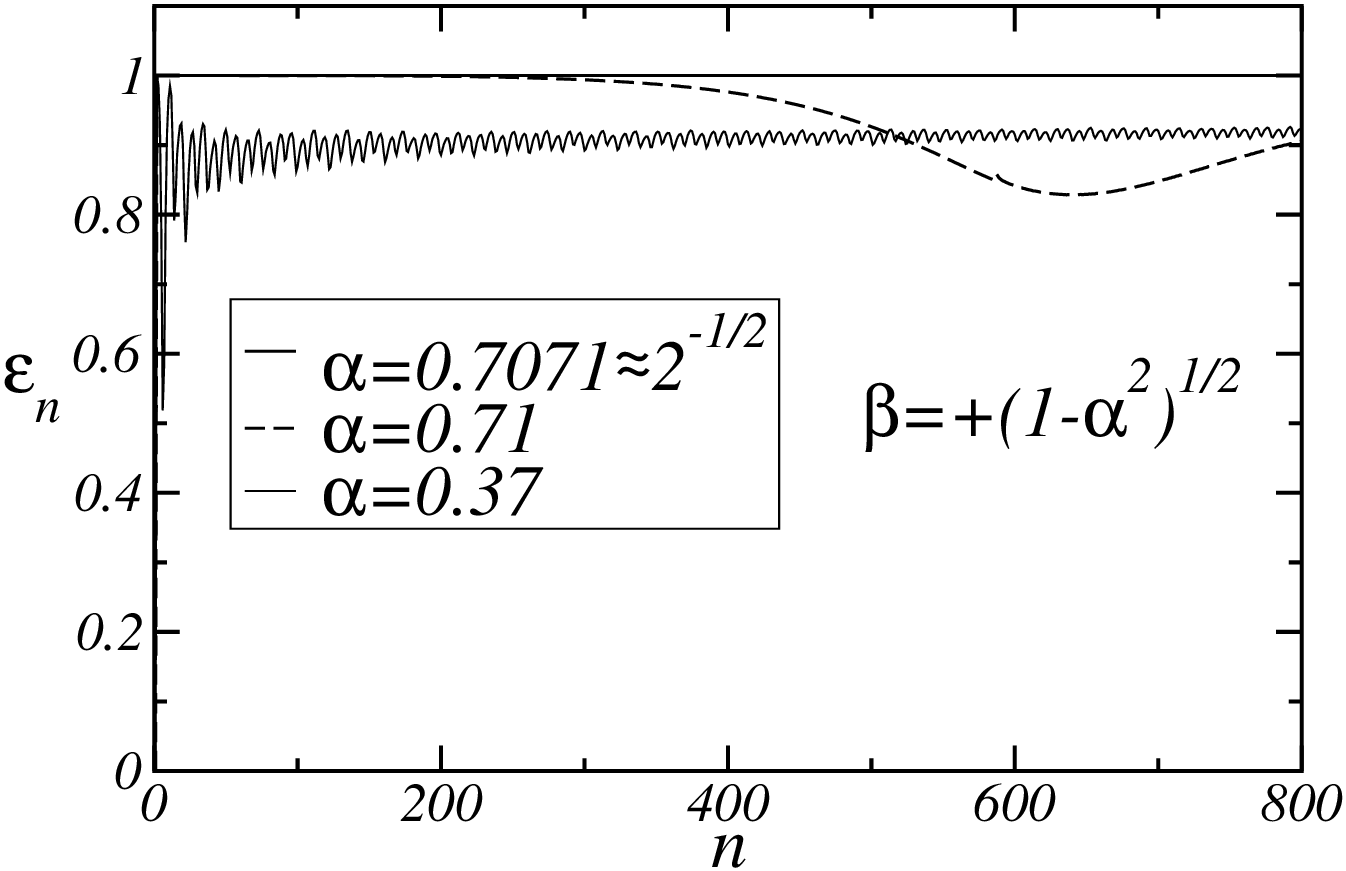}
\caption{Entanglement obtained in the first 800 QRW steps after a down spin measurement with the Hadamard operator
and $\alpha,\beta$ real. The continuous line shows the result for a value of $\alpha$ very close to $\frac{1}{\sqrt{2}}=0.7071067812\cdots$.
The dashed line corresponds to a value of $\alpha$ not so close to $\frac{1}{\sqrt{2}}$ and the lower continuous
(jagged) line to a value decidedly different from $\frac{1}{\sqrt{2}}$.}
\label{Fig2}
\end{figure}

In Fig.~\ref{Fig2} the ratio ${\cal E}_n$ (this time non--averaged) for three different values of $\alpha$ (taking both $\alpha$ and $\beta$
real and positive) are shown as a function of $n$. It is clear that, as $\alpha$ approaches $\beta$, both tending to $\frac{1}{\sqrt{2}}$,
the entanglement gets stuck to the maximum for more and more steps. Therefore we infer from Fig.~\ref{Fig2} that when $\alpha=0.71$
the averaged entanglement after 200 steps is $\overline{\cal E}_{200}\approx1$ while after more steps, say 800, it is patently less than~1,
$\overline{\cal E}_{800}<1$. Instead for $\alpha=0.7071$, a figure closer to $\frac{1}{\sqrt{2}}$, both $\overline{\cal E}_{200}$ and
$\overline{\cal E}_{800}$ are firmly anchored to~1. Recall that, although not shown in this graph, we have proved before that at exactly
$\alpha=\beta=\frac{1}{\sqrt{2}}$ the entanglement drops to zero. The plot for other values of $\alpha$ are qualitatively analogous to
the one shown in the jagged line of Fig.~\ref{Fig2} (here and throughout the rest of the paper, the number 0.37 will just represent an
arbitrary value of $\alpha$). Thus, a method to obtain highly entangled states would be using real values of $\alpha$ and $\beta$, very
close to each other, and look for a down spin measurement output in the first QRW steps. The difficulty is that, as explained above, such
down spin outputs are very unlikely.

\begin{figure}
\includegraphics[scale=0.33]{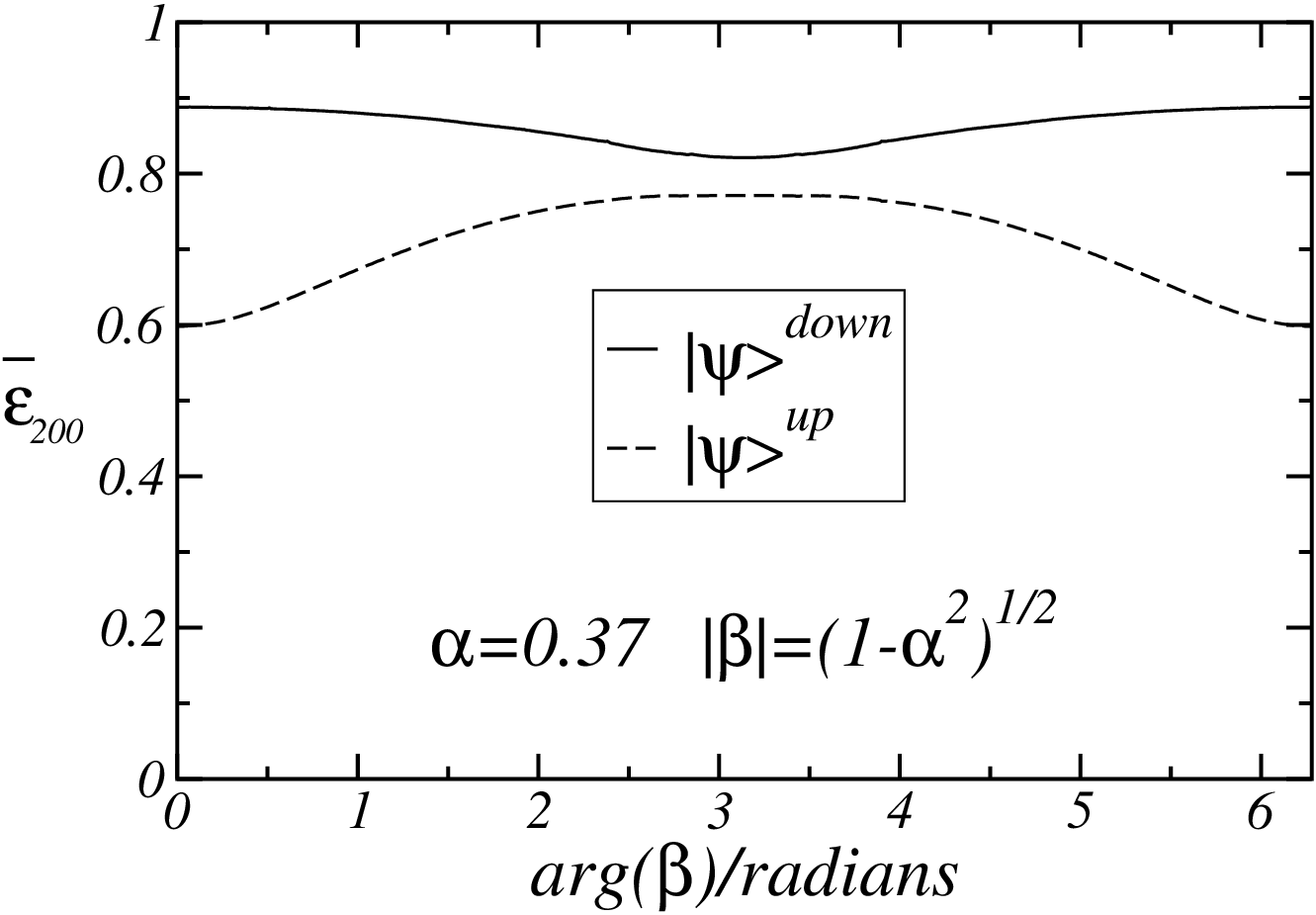}
\caption{Averaged entanglement $\overline{\cal E}_{200}$ as a function of the phase of parameter $\beta$
for a value of $\alpha$ far from $\frac{1}{\sqrt{2}}$ with the Hadamard coin.}
\label{Fig3}
\end{figure}

\begin{figure}
\includegraphics[scale=0.33]{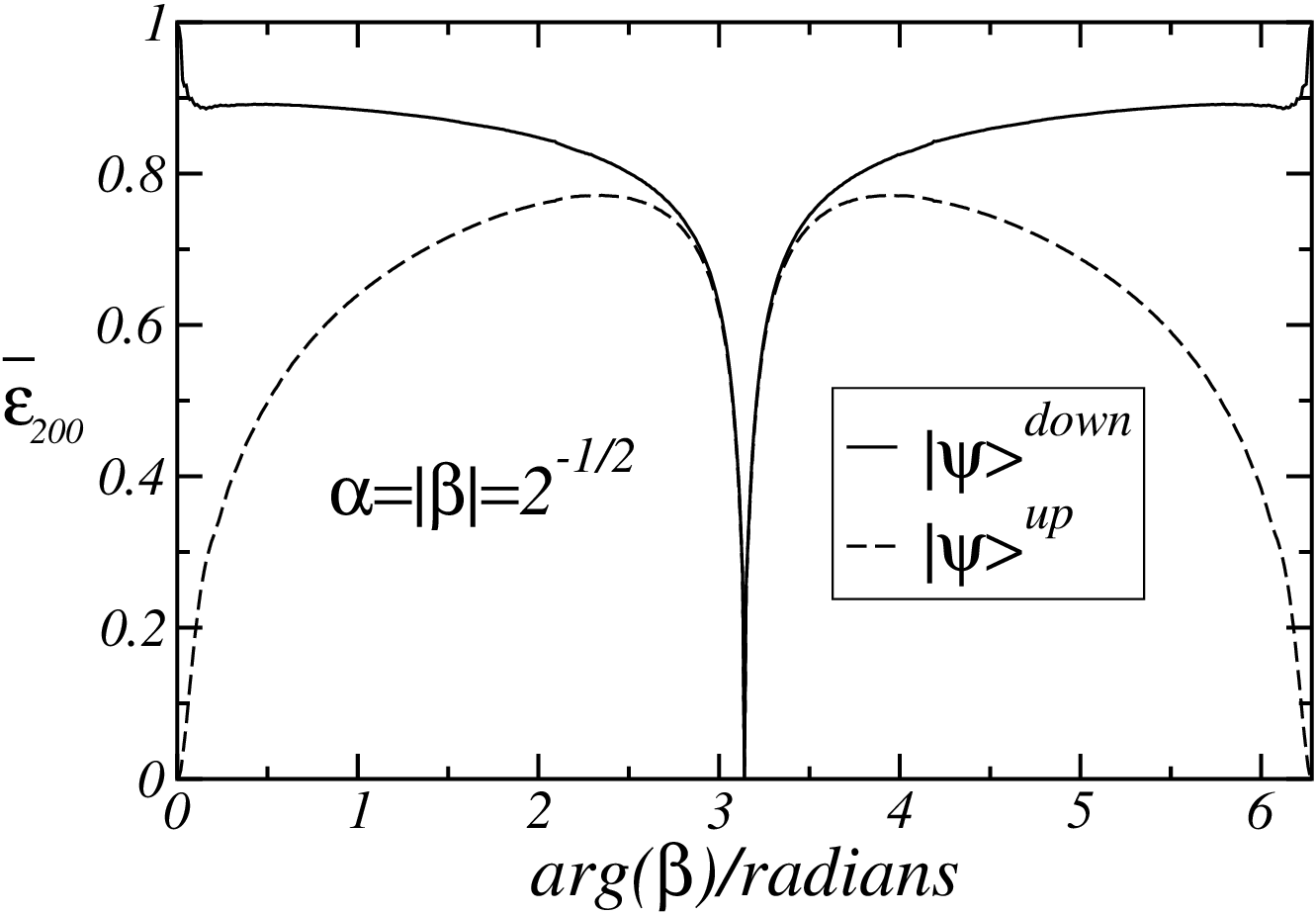}
\caption{Averaged entanglement $\overline{\cal E}_{200}$ as a function of the phase of parameter $\beta$
for $\alpha=\frac{1}{\sqrt{2}}$ with the Hadamard coin.}
\label{Fig4}
\end{figure}

Next we add a phase to $\beta$ while keeping $\alpha$ real. In this way we will have covered all relevant (real or complex) numerical
values for $\alpha$, $\beta$. No dramatic changes in the entanglement generation are found. In particular, no cases with
arg$(\beta)\not=0$ exist with maximal entanglement, see Fig.~\ref{Fig3} and Fig.~\ref{Fig4}. When $\alpha\not=\frac{1}{\sqrt{2}}$
the dependence on arg$(\beta)$ is qualitatively as in Fig.~\ref{Fig3}. For $\alpha=-\beta=\frac{1}{\sqrt{2}}$ the averaged entanglement
goes to zero, as Fig.~\ref{Fig4} shows manifestly, meaning that the true, non--averaged entanglement vanishes at all the QRW steps.

The use of the averaged entanglement $\overline{\cal E}_n$ after $n$ QRW steps must not divert our attention away from the fact
that the real interesting quantity is the normalized entanglement ${\cal E}_n$ achieved by the quantum state after each single
measurement. Our computer code allows to discover isolated cases of maximal entanglement (being isolated, they get lost in the
average of $\overline{\cal E}_n$). They occur mainly in the first few steps of the QRW and are quite likely since the
probability of obtaining the related spin measurement is not very low.

Let us enumerate all cases of this kind found when using the Hadamard coin. We start by the position states originated by
measurements that led to down spin: the entanglement is a maximum at the second step for all $\alpha$ or $\beta$ (real or complex)
(except for $\alpha=\beta=\frac{1}{\sqrt{2}}$ as we already know) and at steps $n=3$ and~4 when the argument of $\beta$ is
arg$(\beta)=\pi/2,3\pi/2$. If instead the output of the measurement at $n=2$ is up spin, then the
entanglement is a maximum for all $\alpha$ real and arg$(\beta)=\pi/2,3\pi/2$. When $n=2$ these features can be proved by
resorting to expressions~(\ref{condition_upstate}) and~(\ref{condition_down}) (but the formalism of section~\ref{section3}
does not suffice to investigate the presence or absence of maximal entanglement beyond the second QRW step). Moreover, in all cases
the probability for obtaining the indicated result from the spin measurement is around $P\sim0.2\, -\, 0.5$ and the number $N$ of
entangled terms in the wavefunction collapsed after the measurement coincides with the QRW step number, $N=n$ except when $n=3$ for
which $N$ is~2. There seems to be no other steps $n$ with ${\cal E}_n$ equal to~1. We have checked this assertion up to $n$ as large as 1000.

\subsection{Kempe coin}
\label{section4.K}

\begin{figure}[t!]
\includegraphics[scale=0.33]{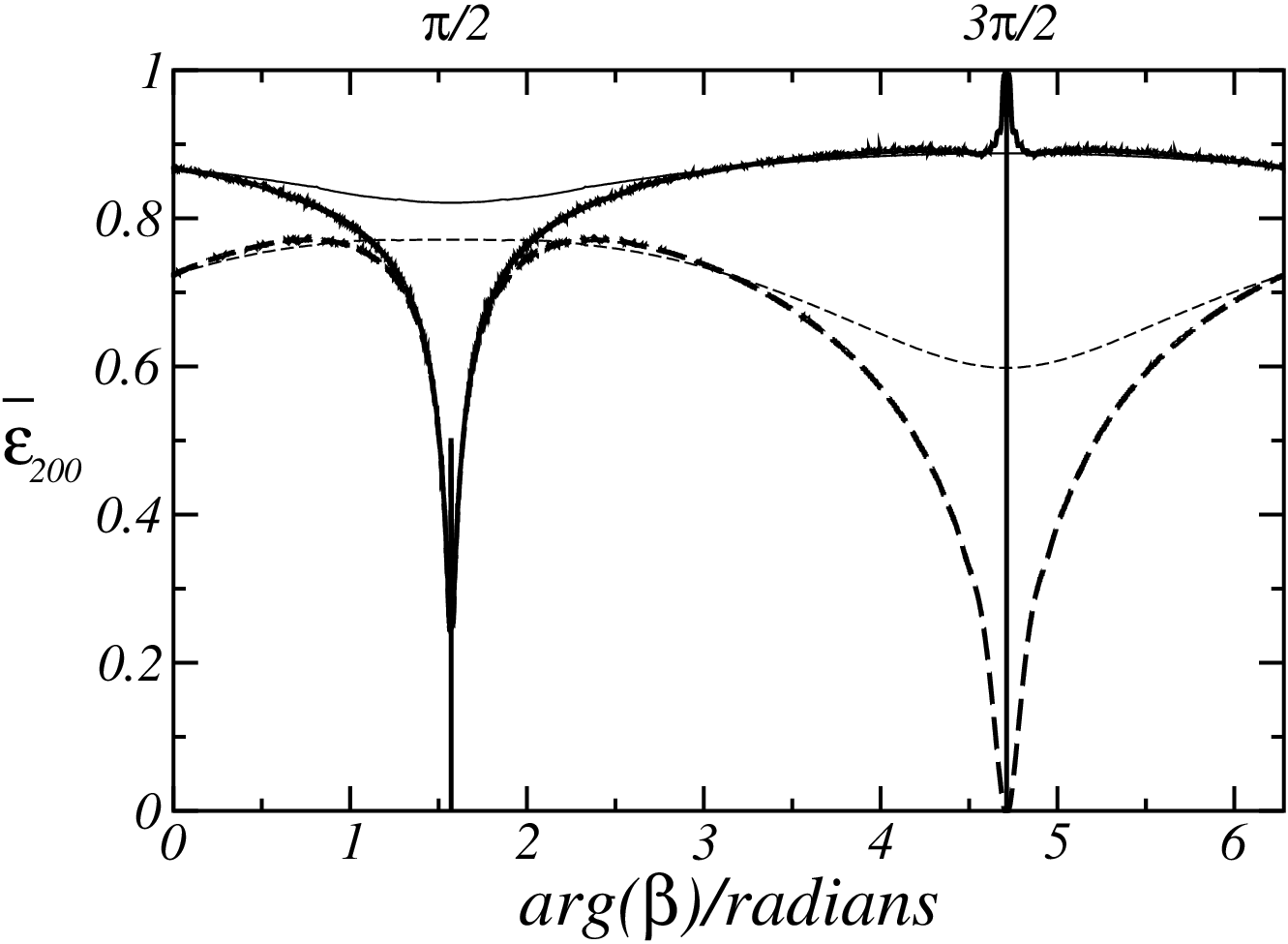}
\caption{Averaged entanglement $\overline{\cal E}_{200}$ as a function of the phase of parameter $\beta$
for various values of $\alpha$ with the Kempe coin. The thick continuous line refers to down spin measurement output
with $\alpha=\vert\beta\vert=\frac{1}{\sqrt{2}}$; the thick dashed line to up spin and
$\alpha=\vert\beta\vert=\frac{1}{\sqrt{2}}$; the thin continuous line to down spin with $\alpha=0.37$
and $\vert\beta\vert=+\sqrt{1-\alpha^2}$; and the thin dashed line to up spin with $\alpha=0.37$ and
$\vert\beta\vert=+\sqrt{1-\alpha^2}$.}
\label{Fig5}
\end{figure}

When using the Kempe coin~(\ref{analytic.kempe}), the average entanglement $\overline{\cal E}_n$ turns out to be independent of
$\alpha$,$\beta$ as far as both coefficients are real and positive
(but depends on $n$ and on the result, up or down spin, of the measurement). However, the dependence
on the argument of a complex $\beta$ displays the rich structure shown in Fig.~\ref{Fig5}. The first aspect of this graph to highlight is
that, once fixed the result of the spin measurement, the averaged entanglements at real and positive $\alpha$,$\beta$
(the rightmost or leftmost edges in the figure) are indeed the same for the two values of $\alpha$ that have
been utilized. The second aspect is that for down spin measurements and $\alpha=\vert\beta\vert=\frac{1}{\sqrt{2}}$
there is maximal entanglement as arg$(\beta)$ tends to become $3\pi/2$ but at this precise value, it suddenly drops
to zero (indicated by the vertical line at this value of arg$(\beta)$), a phenomenon similar to the one described for the Hadamard
coin in Fig.~\ref{Fig1}. In fact, at $\alpha=\frac{1}{\sqrt{2}}$ and $\beta=-\frac{{\rm i}}{\sqrt{2}}$ the wavefunction after $n$
steps is $\vert\phi\rangle_n=\vert\uparrow\rangle\otimes\vert n,n\rangle$. As also happened for real values $\alpha\approx\beta$ in the
Hadamard coin, the probability of having this output after the spin measurement is negligibly small for arg$(\beta)\approx3\pi/2$.

\vskip 2mm

\vskip 2mm

\vskip 2mm

The behavior near arg$(\beta)=\pi/2$ can be described as follows: the averaged entanglement becomes
$\overline{\cal E}_{200}=0.5$ for $\alpha=\vert\beta\vert=\frac{1}{\sqrt{2}}$ and arg$(\beta)$ close to $\pi/2$ while it
drops to zero when the phase attains this precise value. This pattern can be spelled out by studying the non--averaged entanglement:
for $\alpha\approx\vert\beta\vert\approx\frac{1}{\sqrt{2}}$ and arg$(\beta)=\pi/2$ it is maximal for even QRW steps and a
down spin measurement while is zero for odd steps. Instead, for up spin output, the order is reverted: for even steps is zero and
for odd steps is maximal. This explains the value~0.5 that $\overline{\cal E}_{200}$ attains for both spin measurements near
arg$(\beta)=\pi/2$ in Fig.\ref{Fig5} (in other places where $\overline{\cal E}_{200}$
equals~0.5 the above--described alternance between even and odd steps is not seen). Unfortunately the probability of having
such measurement outputs (down spin for even steps or up spin for odd steps) is negligibly small, $P\ll1$. Furthermore, in all
these cases the number of terms in the collapsed wavefunction is just $N=2$ for any step $n$ denoting a rather poor entanglement.
When exactly $\alpha=\vert\beta\vert=\frac{1}{\sqrt{2}}$ and arg$(\beta)=\pi/2$, the entanglement vanishes at all steps for any
measurement output (actually, the wavefunction after $n$ QRW steps is $\vert\phi\rangle_n=\vert\downarrow\rangle\otimes\vert -n,-n\rangle$).
This accounts for the zero of $\overline{\cal E}_{200}$ in Fig.~\ref{Fig5} at strictly arg$(\beta)=\pi/2$
and $\alpha=\vert\beta\vert=\frac{1}{\sqrt{2}}$.

Again we study the non--averaged entanglement during the initial steps fo the QRW.
If the measurement output is down spin then entanglement is maximal at the
steps $n=2,3,4$ of the QRW for all $\alpha$,$\beta$ real and at the step $n=2$ for $\alpha$ real and any complex $\beta$ (the cases
for $n=2$ follow also from~(\ref{condition_down})). When instead the measurement gives up spin, the QRW produces a state with maximal
entanglement only at the second step for all $\alpha$ and $\beta$ real (it matches with the discussion after
expression~(\ref{condition_up})). All the above events with maximal entanglement have a significant probability to occur,
$P\sim0.2\, -\, 0.5$ and the number of terms $N$ in the collapsed position wavefunction coincides with the QRW step $n$ except for
$n=3$, down spin and $\alpha$,$\beta$ real, for which $N$ is~2. Up to $n=1000$ we have found no other maximum non--averaged
entanglement results for any coin measurement output.

\subsection{$Z$ coin}
\label{section4.Z}

\begin{figure}[t!]
\includegraphics[scale=0.33]{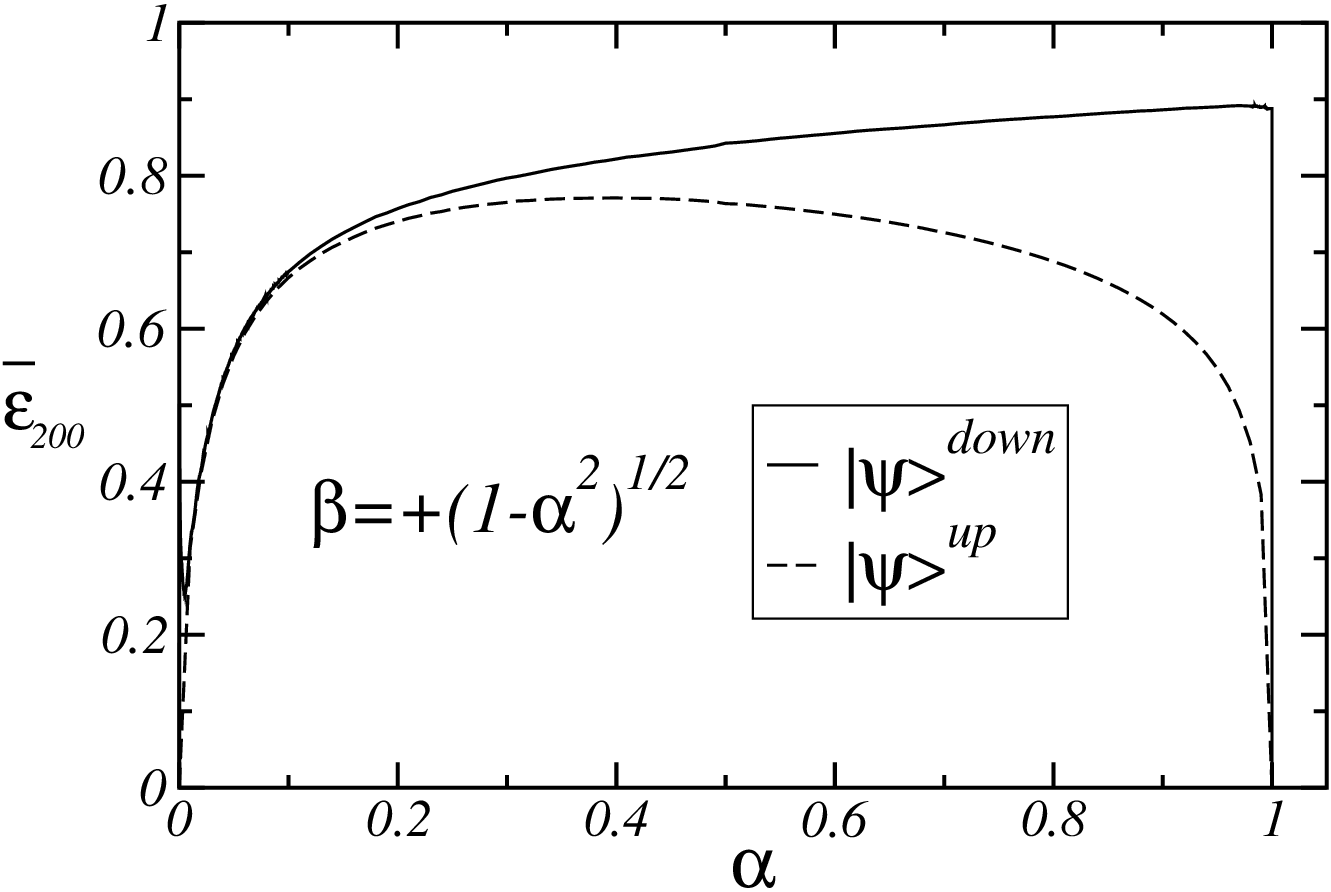}
\caption{Averaged entanglement $\overline{\cal E}_{200}$ as a function of $\alpha$ for real $\alpha$,$\beta$
with the $Z$ coin.}
\label{Fig6}
\end{figure}

In Fig.~\ref{Fig6} we show the averaged entanglement obtained with the use of the $Z$ coin~(\ref{analytic.3}) as a
function of $\alpha$ for real $\beta$. From this plot it is apparent that there are no values of these parameters that
allow to have maximal or almost maximal entanglement during the entire QRW. Moreover, having fixed the (real) values
of $\alpha$ and $\vert\beta\vert$, the plot of
$\overline{\cal E}_{200}$ as a function of arg$(\beta)$ (not shown) comes out flat, indicating no dependence on this phase.
Therefore we conclude that the averaged entanglement never attains the maximum when the $Z$ coin operator is used,
implying that there are no values of $\alpha$,$\beta$ that allow maximal entanglement for all (or almost all) $n>2$.

\vskip 1cm

Maximal entanglement can be achieved only in specific cases. A detailed analysis performed with the non--averaged
entanglement enables us to discover that indeed this event happens for all real $\alpha$ and all complex $\beta$ when the
measurement output yields down spin at the step $n=2$ of the QRW. The number of entangled terms in the collapsed
wavefunction $\vert\psi\rangle^{\rm down}_2$ after the spin measurement is $N=2$ and the probability is $P\sim0.2\,-\,0.5$,
again definitely far from zero. Furthermore, for $\alpha=\vert\beta\vert=\frac{1}{\sqrt{2}}$ and arbitrary arg$(\beta)$
maximal entanglement appears for $n=2,3,4$ steps and with a similarly high probability. In this case, though, the number $N$
of entangled terms in the resulting wavefunction after the spin measurement is equal to the step $n$ only for $n=2,4$
while for $n=3$, $N$ is~2. On the other hand, if the measurement output is up spin, then the
entanglement will become maximal only when $\alpha=\vert\beta\vert=\frac{1}{\sqrt{2}}$ and arbitrary arg$(\beta)$
at the second step with $P=0.5$ and $N=2$ entangled terms. As $\alpha$ and $\vert\beta\vert$ tend to be equal,
the entanglement grows and the probability approaches~0.5.
All these results for $n=2$ can be easily recovered by using the formalism of section~\ref{section3}. In particular, we have
$\vert\psi\rangle^{\rm up}_2\propto\alpha^2\vert2,2\rangle+\vert\beta\vert^2\vert0,0\rangle$ and
$\vert\psi\rangle^{\rm down}_2\propto\alpha^\ast\beta^\ast\vert-2,-2\rangle-\alpha\beta^\ast\vert0,0\rangle$.

\subsection{The general coin}
\label{section4.general}

We have also run a battery of QRW with the general coin operator (expression~(\ref{analytic.1}) with $\varphi=0$) in
order to look for precise combinations of the parameters in $U_{\rm coin}$ allowing maximal entanglement.
The initial state was still given by~(\ref{stateinitial2}).

Two kinds of runs were done. In the first one the averaged entanglement was calculated after 200 QRW steps for
various values of $\rho$, $\theta$, $\eta$ (in $U_{\rm coin}$) and $\alpha$, arg$(\beta)$ (in $U_{\rm shift}$).
The ranges being $0\leq\rho,\alpha\leq1$, $0\leq\theta,\eta\leq\pi$
and $0\leq{\rm arg}(\beta)<2\pi$, data were taken by varying the values of each parameter by jumps of~0.05.
The goal was to find QRW presenting maximal entanglement during all the steps.

The second kind of run was designed to find isolated cases of maximal entanglement during the first few steps of
the QRW. For this reason the QRW were evolved only for 10 steps and the non--averaged normalized entanglement was
extracted. The same five parameters as before were swept within their ranges.

No highly probable parameter regions producing maximal or near maximal averaged entanglement
($\overline{\cal E}_{200}>0.99$ and $P>0.15$) were found. Instead our computer runs were plenty of
isolated cases with maximal entanglement at steps $n=2,3,4$ of the QRW, (${\cal E}_n=1$ and $P>0.15$).
There were so many of them that giving a complete record lies beyond the limits of this short paper.
Most of those at step $n=2$ are summarized in the general discussion for arbitrary $\rho$ after~(\ref{condition_down}).

\section{Conclusions}
\label{section5}

A quantum random walk (QRW) model with two walkers has been devised to study the generation of maximum
entanglement among the two walkers by the process of coin measurement. The model is rather unphysical but
it has been formulated in the way it is in order to maximize the entanglement among walkers.

The allowed positions of the walkers are the set of (positive, negative or null) integers along an infinite line. The 2--state coin
is represented by a spin $\frac{1}{2}$ system. Every QRW was started with the state $\vert\uparrow\rangle\otimes\vert0,0\rangle$,
that is, the spin in the ($Z$--component) up eigenstate and the walkers at the origin of the line. The only consideration
used to select this choice of the initial state was starting the walk with a position state free of entanglement. With this premise,
and by a redefinition of the origin, any initial position eigenstate is as valid as any other. The spin state
is also completely general because by an adequate definition of the axes orientation any spin state can be viewed as the up spin
eigenstate along the $Z$--axis.

We have introduced a shift operator in the QRW with two walkers. It includes two free parameters that,
together with those present in the coin operator, provides enough freedom with which a
measurement performed on the coin state may yield maximal entanglement on the resulting position quantum state. We have
studied the problem both analytically along the first steps of the QRW and by a numerical computer code that
allows to probe an arbitrarily large number of QRW steps. Besides the entanglement, we have measured
the probability of having such highly entangled states and the quality of their entanglement, given by the number
of terms that the collapsed position wavefunction contains. We have also devised an averaged measure of the
entanglement that simplifies the search for a QRW with all or almost all steps presenting high entanglement.

We have used the Hadamard, Kempe and $Z$ coin operators as they are the most generally seen throughout
the literature. Moreover, in subsection~\ref{section4.general} we also tackled the general coin operator.

In general maximal entanglement is more frequently generated when the spin measurement output is opposite to the
one used as initial state, in our case down spin.

We have found two kinds of situations with maximal entanglement: {\it (i)} near maximal entanglement throughout almost the entire QRW,
regardless of the number of steps, but with negligible probability to occur $P\ll1$ (the larger is the number of steps, the lower is
the value of $P$) and {\it (ii)} exactly maximal entanglement during
a few steps at the beginning of the QRW with moderately high probabilities, $P\sim0.2\,-\,0.5$.

There are no highly probable entangled states containing a large number of entangled terms. Only during the first steps of the QRW
one can achieve maximal entanglement while having a reasonably high probability. However, such states contain a limited number
of entangled terms precisely because they occur at the first steps of the QRW. This happens for instance for down spin measurement at the
second step for arbitrary $\alpha$ and arg$(\beta)$ with any of the three coins mentioned in section~\ref{section2}.
The largest absolute entanglement with an acceptably high probability and with the above three coins is achieved after a down
spin measurement, occurring with probability $P=0.25$, at the $n=4$ step and managing to entangle $N=4$ terms in the position wavefunction
(thus giving $E_4=\log_24=2$). They are: {\it (i)} Hadamard coin for all $\alpha$ and arg$(\beta)=\pi/2,3\pi/2$, {\it (ii)}
Kempe coin with any $\alpha$,$\beta$ real and {\it (iii)} $Z$ coin with $\alpha=\vert\beta\vert=\frac{1}{\sqrt{2}}$ and all arg$(\beta)$.

\section{Acknowledgements}

Part of the work was done during a visit of B.A. to the Hacettepe University. B.A. thanks the warm hospitality during
the stay. B.A. also wishes to thank Emilio d'Emilio and Hans--Thomas Elze for useful conversations and Vivien Kendon for
providing us with some references on experimental realizations of QRW.
We acknowledge the CINECA (Bologna, Italy) ISCRA Award N. HP10CB7JY6.2010 for the availability of high performance
computing resources and support.


\end{multicols}
\end{document}